\documentclass[journal]{IEEEtran}
\usepackage{graphicx,amssymb,amsmath}
\usepackage{multicol}
\usepackage[noadjust]{cite}
\usepackage{setspace}
\usepackage{stfloats}
\usepackage{midfloat}
\usepackage[normal]{threeparttable}
\usepackage{flushend,cuted}
\usepackage[normal]{threeparttable}
\usepackage{balance}
\usepackage{cuted}
\usepackage{cases,subeqnarray}
\usepackage{bm,multirow,bigstrut}
\usepackage{textcomp}
\usepackage{latexsym,bm}
\usepackage{booktabs,changebar}
\usepackage{xcolor}
\usepackage{mathtools}
\newtheorem{lemm}{Lemma}

\newtheorem{coro}{Corollary}

\IEEEoverridecommandlockouts
\begin{document}

\title{Spectral Efficiency of Multipair Massive MIMO Two-Way Relaying with Hardware Impairments}
%



\author{Jiayi Zhang,~\IEEEmembership{Member,~IEEE}, Xipeng Xue, Emil~Bj\"{o}rnson,~\IEEEmembership{Member,~IEEE}, Bo Ai, and Shi Jin,~\IEEEmembership{Member,~IEEE}
\thanks{This work was supported in part by the National Natural Science Foundation of China (Grant Nos. 61601020), and the Fundamental Research Funds for the Central Universities (Grant Nos. 2016RC013 and 2016JBZ003). This work is also supported by ELLIIT and CENIIT. The work of S. Jin was supported in part by the National Science Foundation (NSFC) for Distinguished Young Scholars of China with Grant 61625106.}%
\thanks{J. Zhang and X. Xue are with the School of Electronic and Information Engineering, Beijing Jiaotong University, Beijing 100044, P. R. China (e-mail: jiayizhang@bjtu.edu.cn).}
\thanks{E. Bj\"{o}rnson is with the Department of Electrical Engineering (ISY), Link\"{o}ping University, Link\"{o}ping, Sweden.}
\thanks{Bo Ai is with the State Key Laboratory of Rail Traffic Control and Safety, Beijing Jiaotong University, Beijing 100044, P. R. China.}
\thanks{S. Jin is with the National Mobile Communications Research Laboratory, Southeast University, Nanjing 210096, P. R. China.}
}
\maketitle
\begin{abstract}
We consider a multipair massive multiple-input multiple-output (MIMO) two-way relaying system, where multiple pairs of single-antenna devices exchange data with the help of a relay employing a large number of antennas $N$. The relay consists of low-cost components that suffer from hardware impairments. A large-scale approximation of the spectral efficiency (SE) with maximum ratio (MR) processing is derived in closed form, and the approximation is tight as $N \to \infty$. It is revealed that for a fixed hardware quality, the impact of the hardware impairments vanishes asymptotically when $N$ grows large. Moreover, the impact of the impairments may even vanish when the hardware quality is gradually decreased with $N$, if a scaling law is satisfied. Finally, numerical results validate that multipair massive MIMO two-way relaying systems are robust to hardware impairments at the relay.
\end{abstract}


\section{Introduction}
Massive multiple-input multiple-output (MIMO) is a promising technology for future wireless cellular networks. By equipping base stations (BSs) with hundreds of antenna elements, the spectral efficiency (SE)
of massive MIMO systems can be improved by several orders of magnitude compared to conventional small-scale MIMO systems \cite{ngo2013energy}.
On a parallel avenue, the two-way relaying technology has been developed to allow a user pair to exchange information through an intermediate relay. Thanks to the simple signal processing when using arrays with many antennas, the combination of massive MIMO and two-way relaying is an attractive candidate for future wireless systems \cite{wong2017key,hongbo2016ubiquitous}.

Motivated by these observations, some researchers have analyzed the performance of multipair massive MIMO two-way relaying systems. In \cite{cui2014multi}, the asymptotic SE of the multipair two-way relaying system was obtained analytically by using a relay with very many antennas and maximum ratio (MR) processing. It was revealed that the SE increases logarithmically with the number of antennas at the relay, but decreases logarithmically with the number of user pairs \cite{jin2015ergodic}. For the multipair massive MIMO one-way full-duplex relaying system, an achievable rate expression in closed form for MR processing and an analytical approximation of the achievable rate for zero forcing processing were derived in \cite{ngo2014multipair}. {\cite{zhang2016spectral,Sharma2017Energy} quantified the asymptotic SE and energy efficiency for multipair massive MIMO two-way full-duplex relay systems.}

Different from most of the existing works, which consider systems with ideal hardware, herein we consider a massive MIMO two-way relaying system with transceiver hardware impairments at the relay. This is motivated by the fact that cost-efficient deployment of arrays with many antennas require low-quality components. If the same hardware quality is used in massive MIMO systems as in conventional systems \cite{li2010cooperative}, then the hardware cost would scale linearly with the number of antennas \cite{zhang2016achievable}. An alternative low-cost implementation is one with components of reduced quality, but this leads to non-negligible distortion from hardware impairments. {In practical systems, many transceiver components act as non-linear filters that induce hardware impairments. The transceiver hardware impairments include amplifier non-linearities, phase noise, I/Q imbalance, and quantization errors \cite{zhang2016on,zhang2017performance,bjornson2014massive}. In practice, these impairments can be mitigated at the relay by using appropriate compensation algorithms \cite{schenk2008rf}, but residual impairments always remain. We assume that appropriate compensation algorithms have been applied and focus on the residual hardware impairments. As shown in \cite{schenk2008rf}, the residual hardware impairments at the transmitter and receiver can be modeled as additive independent distortion terms that are proportional to the signal power at the transmitter and receiver, respectively. Since the received signal power depends on the current channel realizations, so does the variance of the receiver distortion.}
In this letter, we try to answer the question whether the high SE of multipair massive MIMO two-way relaying can be achieved also under residual hardware impairments.

\section{System Model}\label{se:system}

Let us consider a two-way system model where $K$ pairs of single-antenna devices communicate with each other through an $N$-antennas relay ${T_R}$ \cite{jin2015ergodic}. The devices, which are denoted as
$ {{T_{{A_i}}}} $ and ${{T_{{B_i}}}} $, for $i = 1,\ldots,K $, could, for example, be user terminals that exchange information or small-cell BSs that need a backhaul link. We assume that ${T_{{A_i}}}$ and ${T_{{B_i}}}$ cannot exchange information directly due to large geometric path loss and/or heavy shadowing.

The relay and all devices operate in time-division-duplex mode so that channels between each device and relay are reciprocal within every channel coherence block. We further model each channel as an ergodic stationary process with a fixed independent realization in each coherence block. We define ${\bf{G}} \buildrel \Delta \over = \left[ {{{\bf{g}}_1},\ldots,{{\bf{g}}_K}} \right]$ and ${\bf{H}} \buildrel \Delta \over = \left[ {{{\bf{h}}_1},\ldots,{{\bf{h}}_K}} \right]$, where ${{\bf{g}}_i} \in {\mathbb{C}^{N \times 1}}$ and ${{\bf{h}}_i}\in {\mathbb{C}^{N \times 1}} $, for $i = 1,\ldots,K $, denote the uplink channel vector between ${T_{{A_i}}}$ and ${T_R}$ and the downlink channel vector between ${T_{{B_i}}}$ and ${T_R}$, respectively. Furthermore, $\bf{G}$ and $\bf{H}$ can be expressed as ${\bf{G}} = {{\bf{S}}_u}{\bf{D}}_u^{1/2}$ and ${\bf{H}} = {{\bf{S}}_d}{\bf{D}}_d^{1/2}$, respectively. {Motivated by the channel measurements in \cite{schenk2008rf,Gao2014Massive}, the elements of ${{\bf{S}}_u}$ and ${{\bf{S}}_d}$ are assumed to be independent and identical distributed (i.i.d.) as ${\cal{CN}}({ {0}}, 1)$.} ${{\bf{D}}_u}$ and ${{\bf{D}}_d}$ are diagonal matrices representing the large-scale fading and the $i$th diagonal elements of ${{\bf{D}}_u}$ and ${{\bf{D}}_d}$ are denoted as $\sigma _{{g_i}}^2$ and $\sigma _{{h_i}}^2$, respectively. Moreover, the columns of $\bf{G}$ and $\bf{H}$, i.e., ${\bf{g}}_i$ and ${\bf{h}}_i$ are i.i.d. as ${\cal{CN}}({\bf{0}}, \sigma _{{g_i}}^2{{\bf{I}}_N})$ and ${\cal{CN}}({\bf{0}}, \sigma _{{h_i}}^2{{\bf{I}}_N})$, respectively. By exploiting the channel reciprocity, the downlink channel between ${T_{{A_i}}}$ and ${T_R}$ and between ${T_{{B_i}}}$ and ${T_R}$ are denoted by ${\bf{g}}_i^T$ and ${\bf{h}}_i^T$, respectively.

The communication in the multipair two-way relaying system consists of two phases. In the first phase, all devices ${T_{{A_i}}}$ and ${T_{{B_i}}}$ transmit their own information-bearing signals, i.e., ${x_{{A_i}}}$ and ${x_{{B_i}}}$ to relay ${T_R}$, respectively. We assume that ${x_{{A_i}}}$ and ${x_{{B_i}}}$ are Gaussian data signals, and all devices have the same transmit power, denoted by ${P_U}$. The additive distortion  ${\boldsymbol{\eta} _r} $ describes the impairments of the receiver hardware at the relay\footnote{{It is well-known that the impact of hardware impairments in the single-antenna terminals is the same as in single-antenna systems. Therefore, we only consider the hardware impairments at the multi-antenna relay.}}. {${\boldsymbol{\eta} _r} $ is proportional to the instantaneous received signal power at the antenna as ${\boldsymbol{\eta} _r} \sim {\cal{CN}}({\bf{0}},\kappa _r^2{P_U} {\text{diag}} ({{W}}_{11}, \dots,{{W}}_{NN} ))$, where $W_{ii}$ is the $i$th diagonal element of ${\bf{W}} = \sum_{j=1}^{K} ({\bf{h}}_j {\bf{h}}_j^H +{\bf{g}}_j {\bf{g}}_j^H) $ \cite{zhang2016achievable,bjornson2014massive}.} The proportionality coefficient $\kappa _r$ characterizes the level of receiver hardware impairments. Then, the received signal at ${T_R}$ can be expressed as
\begin{align}\label{receive signals at relay}
{{\bf{y}}_r} = \sum\limits_{j = 1}^K {({{\bf{g}}_i}{x_{{A_i}}} + {{\bf{h}}_i}{x_{{B_i}}})}  + {\boldsymbol{\eta} _r} + {{\bf{n}}_R},
\end{align}
where ${{\bf{n}}_R} \sim {\cal{CN}}({\bf{0}},\sigma _R^2{{\bf{I}}_N})$ is the additive white Gaussian noise (AWGN) at ${T_R}$. We can rewrite \eqref{receive signals at relay} in matrix form as
\begin{align}\label{another expression of receive signals at relay}
{{\bf{y}}_r} = {\bf{A}}{\bf{x}} + {\boldsymbol{\eta} _r} + {{\bf{n}}_R},
\end{align}
where ${\bf{A}} \buildrel \Delta \over = \left[ {{\bf{G}},{\bf{H}}} \right]$, ${\bf{x}} \buildrel \Delta \over = {\left[ {{\bf{x}}_A^T,{\bf{x}}_B^T} \right]^T}$, ${{\bf{x}}_A} \buildrel \Delta \over = \left[ {{x_{{A_1}}},\ldots,{x_{{A_K}}}} \right]^T$, and ${{\bf{x}}_B} \buildrel \Delta \over = \left[ {{x_{{B_1}}},\ldots,{x_{{B_K}}}} \right]^T$.

In the second phase, ${T_R}$ multiplies ${{\bf{y}}_r}$ with the precoding matrix ${\bf{F}} \in {\mathbb{C}^{N \times N}} $ and a power-control coefficient ${\rho}$\footnote{{Due to its low complexity and short latency, the amplify-and-forward scheme is adopted at the relay.}}. Then ${T_R}$ broadcasts ${{{\bf{y}}_t^\prime}}=\rho{\bf{F}}{{\bf{y}}_r} $ to all devices. However, due to the hardware impairment at the transmitter, ${T_R}$ actually broadcasts ${{\bf{y}}_t}$ to all devices as
\begin{align}\label{receive signals at small cell}
{{\bf{y}}_t} = {{\bf{y}}_t^\prime}  + {\boldsymbol{\eta} _t} = \rho {\bf{F}}{{\bf{y}}_r} + {\boldsymbol{\eta} _t},
\end{align}
{where ${\boldsymbol{\eta} _t} \sim {\cal{CN}}\left({\bf{0}},\kappa _t^2\frac{{P_R}}{N} {{\bf{I}}_N}\right)$ with ${P_R}$ being the transmit power of the relay and $\kappa _t$ is the proportionality coefficient of the transmitter hardware impairments \cite{zhang2016achievable,bjornson2014massive}.} Note that $\kappa _r$ and $\kappa _t$ can be interpreted as the error vector magnitude (EVM), which is defined as the ratio of distortion to signal magnitudes. For instance, 3GPP LTE only supports EVMs smaller than $\kappa _t^2 =0.175$ \cite{bjornson2014massive}. For notational convenience, we drop the subscript of hardware impairments at both transmit and receive antennas as $\kappa$. Moreover, the relay can obtain CSI from uplink pilots sent by the devices, and the devices can then obtain CSI through beamforming training \cite{ngo2013energy}. To focus on the effect of hardware impairments, we assume that ${T_R}$ has perfect global CSI, i.e., $\left\{ {{{\bf{g}}_i},{{\bf{h}}_i}} \right\},\forall i$. The power-control coefficient ${\rho}$ is normalized by the instantaneous received signal power
\begin{align}\label{rho}
\rho  = \sqrt {\frac{{{P_R}}}{{{P_U}  {{{\left\| {{\bf{FA}}} \right\|}^2}}   + {{{\left\| {{\bf{F}}{{\boldsymbol{\eta}} _r}} \right\|}^2}}  + {\sigma _R^2} {{{\left\| {\bf{F}} \right\|}^2}}  }}}.
\end{align}
We consider the MR scheme at the relay since it is a low-complexity signal processing algorithm suitable for low-cost massive MIMO deployment \cite{ngo2013energy}. Therefore, the precoding matrix is selected as ${\bf{F}} = {{\bf{B}}^*}{{\bf{A}}^H}$, where ${\bf{B}} = \left[ {{\bf{H}},{\bf{G}}} \right]$. Then, the received signals at ${T_{{A_i}}}$ and ${T_{{B_i}}}$ are
\begin{align}\label{expression of zai}
{Z_{{A_i}}} &= {\bf{g}}_i^T{{\bf{y}}_t} + {n_{{A_i}}}  ,
\end{align}
where ${n_{{A_i}}} \sim {\cal{CN}}(0,\sigma _{{A_i}}^2)$ is AWGN. In the following, we only provide analytical results for ${T_{{A_i}}}$. The SE of ${T_{{B_i}}}$ can be simply derived by switching $A$ and $B$, and $\bf{h}$ and $\bf{g}$. Substituting \eqref{another expression of receive signals at relay} and \eqref{receive signals at small cell} into \eqref{expression of zai}, ${Z_{{A_i}}}$ can be expressed as
\begin{align}\label{expression of expanded zai}
&{Z_{{A_i}}}= \underbrace {\rho {\bf{g}}_i^T{\bf{F}}{{\bf{h}}_i}{x_{{B_i}}}}_{\text{signal}}  + \underbrace {\rho \sum\limits_{j = 1,j \ne i}^K {\left( {{\bf{g}}_i^T{\bf{F}}{{\bf{g}}_j}{x_{{A_j}}} + {\bf{g}}_i^T{\bf{F}}{{\bf{h}}_j}{x_{{B_j}}}} \right)} }_{\text{inter-user interference}} \notag \\
&+ \underbrace {\rho {\bf{g}}_i^T{\bf{F}}{{\bf{g}}_i}{x_{{A_i}}}}_{\text{self-interference}}+ \underbrace {\rho {\bf{g}}_i^T{\bf{F}}{\boldsymbol{\eta} _r} + {\bf{g}}_i^T{\boldsymbol{\eta} _t}}_{\text{hardware impairments}}
  + \underbrace {\rho {\bf{g}}_i^T{\bf{F}}{n_R} + {n_{{A_i}}}}_{\text{compound noise}}.
\end{align}

From \eqref{expression of expanded zai}, we notice that ${Z_{{A_i}}}$ consists of five parts: the signal that ${T_{{A_i}}}$ desires to receive, the inter-user interference, the interference caused by itself, the distortion induced by hardware impairments at relay and the compound noise. With local CSI obtained through beamformed pilots, devices can take advantage of self-interference cancelation to eliminate self-interference \cite{ngo2013energy}. By treating the remaining uncorrelated interference terms as the worst-case independent Gaussian noise in the signal detection, the SE of ${T_{{A_i}}}$ is given by
\begin{align}\label{ergodic achievable rate}
{R_{{A_i}}} = {\mathbb{E}}\left\{ {{{\log }_2}\left( {1 + {\text{SINR}}_{{A_i}}} \right)} \right\},\;\;\; {\text{for}} \ i = 1,\ldots,K,
\end{align}
where ${\mathbb{E}}\left\{ \cdot \right\}$ is the expectation operator and the signal-to-interference plus noise ratio (SINR) of ${T_{{A_i}}}$ is given by
\begin{align}\label{SINR}
{\text{SINR}}_{{A_i}} = \frac{{{P_U}{{\left| {{\bf{g}}_i^T{\bf{F}}{{\bf{h}}_i}} \right|}^2}}}{{C' \!+\! D' \!+\! E'  \!+\! {{\left|  {\bf{g}}_i^T{\bf{F}}{\boldsymbol{\eta} _r} \right|}^2}  \!+\! \frac{1}{{{\rho ^2}}} { {{\left| {\bf{g}}_i^T{\boldsymbol{\eta} _t} \right|}^2}}}},
\end{align}
where $C' \buildrel \Delta \over = {P_U}\sum\limits_{j = 1,j \ne i}^K {\left( {{{\left| {{\bf{g}}_i^T{\bf{F}}{{\bf{g}}_j}} \right|}^2} + {{\left| {{\bf{g}}_i^T{\bf{F}}{{\bf{h}}_j}} \right|}^2}} \right)} $, $D' \buildrel \Delta \over = \sigma _R^2{\left\| {{\bf{g}}_i^T{\bf{F}}} \right\|^2}$, and $E' \buildrel \Delta \over = \frac{{\sigma _{Ai}^2}}{{{\rho ^2}}}$, respectively.

\section{Spectral Efficiency Analysis}\label{analysis}
To the best of the authors' knowledge, finding a closed-form expression for \eqref{ergodic achievable rate} is extremely difficult if not impossible \cite{jin2015ergodic}. In this section, we obtain a closed-form large-scale approximation, which is tight as $N \to \infty$. According to Jensen's inequality, a lower bound on ${R_{{A_i}}}$ is first obtained as
\begin{align}\label{ergodic achievable rate low bound}
{R_{{A_i}}} \ge  {  R_{{A_i}}^{'}} = {\log _2}\left( {1 + \frac{1}{{\mathbb{E}}\left\{ {{{\left[ {{\text{SINR}}_{{A_i}}} \right]}^{ - 1}}} \right\}}} \right).
\end{align}
\begin{lemm}\label{lemm1}
With MR processing and hardware impairments at the relay, the lower bound \eqref{ergodic achievable rate low bound} is given by
\begin{align}\label{An approximate ergodic achievable rate}
{  R_{{A_i}}^{'}} \!-\! \frac{1}{2}{\log _2}\left( {1 \!+\! \frac{N}{{{C_i} \!+\! {D_i} \!+\! {E_i} \!+\!  {F_i} \!+\! {G_i}}}} \right) \!\xrightarrow[{N \to \infty }]{}\! 0,
\end{align}
where ${C_i} \!\buildrel \Delta \over =\! \sum\limits_{j \ne i}^K {\left( {\frac{{\sigma _{{h_j}}^2}}{{\sigma _{{h_i}}^2}} \!+\! \frac{{\sigma_{{h_j}}^4\sigma_{{g_j}}^2}}{{\sigma_{{h_i}}^4\sigma_{{g_i}}^2}} \!+\! \frac{{\sigma _{{g_j}}^2}}{{\sigma_{{h_i}}^2}} \!+\! \frac{{\sigma_{{g_j}}^4\sigma_{{h_j}}^2}}{{\sigma_{{h_i}}^4\sigma_{{g_i}}^2}}} \right)}$, ${D_i} \buildrel \Delta \over = \frac{{{\sigma_R^2}}}{{{P_U}\sigma_{{h_i}}^2}}$, ${E_i} \buildrel \Delta \over= \frac{{{\sigma _{{A_i}}^2}}J}{{{P_R}{\sigma_{{g_i}}^4\sigma_{{h_i}}^4}}}$, ${F_i} \buildrel \Delta \over= {\frac{{\kappa ^2}J}{{\sigma_{{g_i}}^2\sigma_{{h_i}}^4}}}$, ${G_i} = \frac{{{\kappa ^2}}}{{\sigma _{{h_i}}^2}}\sum\limits_{j = 1}^K {\left( {\sigma_{{g_{ij}}}^2 + \sigma_{{h_{ij}}}^2} \right)} $, and \\
$J \!\buildrel \Delta \over=\!  {\sum\limits_{j = 1}^K { {\sigma_{{g_j}}^2\sigma_{{h_j}}^2\left( {\sigma_{{g_j}}^2 \!+\! \sigma_{{h_j}}^2} \right)} } \! +\! \frac{2{\kappa ^2}}{N}\sum\limits_{j = 1}^K { {\sigma_{{g_j}}^2\sigma_{{h_j}}^2} } \sum\limits_{j = 1}^K{\left( {\sigma_{{g_j}}^2 \!+\! \sigma_{{h_j}}^2} \right)} }$.
\end{lemm}

\begin{IEEEproof}
Please refer to the appendix.
\end{IEEEproof}
Lemma \ref{lemm1} reveals that the SE increases when $N$ increases. Further insights can be obtained by investigating the interference terms. First, the SE increases with ${\sigma_{{g_i}}^2}$ and ${\sigma_{{h_i}}^2}$, which means the channel quality for the $i$th device pair is better. On the other hand, ${R_{{A_i}}^{'}}$ decreases if increasing ${\sigma_{{g_j}}^2}$ and ${\sigma_{{h_j}}^2}$, for $j \ne i$, which means the channel conditions for other device pairs except the $i$th device pair is improved. This result is consistent with the results in \cite{jin2015ergodic}. Second, we observe that the compound noise term ${D_i}$ is the inverse of the average SNR for the channel from $T_{B_i}$ to ${T_R}$. Therefore, we can increase the SNR for ${T_{{B_i}}}$ to increase the SE. Furthermore, $ {E_i}$ is composed of the transmit power of ${T_R}$, the pathloss from $T_R$ to ${T_{{A_i}}}$ and the effect of hardware impairments. From $ {E_i}$, it is obvious that the SE will increase when the transmit power of ${T_R}$ increases, but decrease when $\kappa $ increases. Finally, it is clear that the detrimental impact of hardware impairments in ${F_i}$ and ${G_i}$ is independent of $N$. The impact of hardware impairments on the two-way relay system is almost twice as much as that of multiuser MIMO systems \cite{bjornson2014massive}. The two-way relay system suffers from hardware impairments of transceivers, i.e., ${F_i}$ and ${G_i}$, at the relay in the two phases.
\begin{coro}\label{coro:1}
Suppose the hardware impairments level is replaced as $\kappa^2 \triangleq \kappa_0^2 N^z$ for a given scaling exponent $ 0 < z\leq 1$ and an initial value $\kappa_0 > 0$. As $N \to \infty$, the SE converges as
\begin{align}
\left\{ \begin{array}{l}
{R_{{\rm{A}}_i}^{'}} \!-\! \frac{1}{2}{\log _2}\left( {1 \!+\! \frac{{\sigma_{{{\rm{g}}_i}}^2\sigma_{{{\rm{h}}_i}}^4{N^{1 - z}}}}{{\kappa _0^2\left( {{\mu _0}\sigma_{{{\rm{g}}_i}}^2\sigma_{{{\rm{h}}_i}}^2 + {\mu _1}} \right)}}} \right)\xrightarrow[{N \to \infty }]{}  0,\;0 < z < 1\\
{R_{{\rm{A}}_i}^{'}} \!-\! \frac{1}{2}{\log _2}\left( {1 \!+\! \frac{{\sigma_{{{\rm{g}}_i}}^2\sigma_{{{\rm{h}}_i}}^4}}{{\kappa _0^2{{\left( {{\mu _0}\sigma_{{{\rm{g}}_i}}^2\sigma_{{{\rm{h}}_i}}^2 \!+\! {\mu _1} \!+\! 2\kappa _0^{\rm{2}}{\mu _2}} \right)}_i}}}} \right)\xrightarrow[{N \to \infty }]{}  0,z = 1
\end{array} \right.\notag
\end{align}
where ${\mu _0} \buildrel \Delta \over = \sum\limits_{j = 1}^K {\left( {\sigma_{{{\rm{h}}_j}}^2\!+\! \sigma_{{{\rm{g}}_j}}^2} \right)}$, ${\mu _1} \buildrel \Delta \over = \sum\limits_{j = 1}^K {\left( {\sigma_{{{\rm{g}}_j}}^2\sigma _{{{\rm{h}}_j}}^2\left( {\sigma_{{{\rm{g}}_j}}^2 \!+\!  \sigma_{{{\rm{h}}_j}}^2} \right)} \right)} $ and ${\mu _2} \buildrel \Delta \over = \sum\limits_{j = 1}^K {\left( {\sigma_{{{\rm{g}}_j}}^2\sigma_{{{\rm{h}}_j}}^2} \right)} \sum\limits_{j = 1}^K {\left( {\sigma_{{{\rm{g}}_j}}^2 \!+\!  \sigma_{{{\rm{h}}_j}}^2} \right)}$.
\end{coro}
\begin{IEEEproof}
We first substitute $\kappa^2 \triangleq \kappa_0^2 N^z$ into \eqref{An approximate ergodic achievable rate}. As $N \to \infty$ and $z>0$, both ${C_i}$ and ${D_i}$ go to zero, ${E_i}$ behaves as $\mathcal{O}(N^{z-1})$, ${F_i}$ behaves as $\mathcal{O}(N^{z}+N^{2z-1})$, while ${G_i}$ behaves as $\mathcal{O}(N^z)$. To make the numerator and denominator have the same scaling and be non-vanishing, we need to fulfill $1-\max(z,2z-1) \geq 0$ as $ 0 < z\leq 1$.
\end{IEEEproof}

The hardware scaling law in Corollary \ref{coro:1} is different from the one for multiuser massive MIMO systems in \cite{bjornson2014massive}. It proves that larger hardware impairments can be tolerated with increasing number of antennas at the relay in multipair massive MIMO two-way relaying systems. Recall that the EVM at the relay is defined as ${\text{EVM}} = \kappa$ \cite{bjornson2014massive}, which means it can be increased proportionally to $N^{1/2}$. For example, a high-quality relay antenna unit with an EVM of 0.03 can thus be replaced by 16 low-quality antenna units with an EVM of 0.12, while the loss in SE is negligible. Therefore, having a large number of antennas allows the relay to use low-quality hardware.

\section{Numerical Results}\label{numerical}

\begin{figure}[t]
\centering
\includegraphics[scale=0.55]{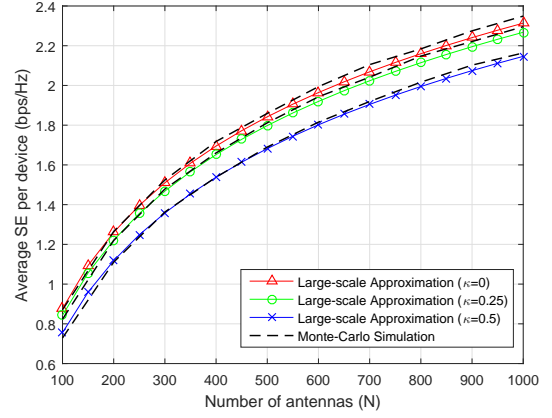}
\caption{Average SE per device for different values of $\kappa$.}
\label{Fig.3}
\end{figure}

\begin{figure}[t]
\centering
\includegraphics[scale=0.55]{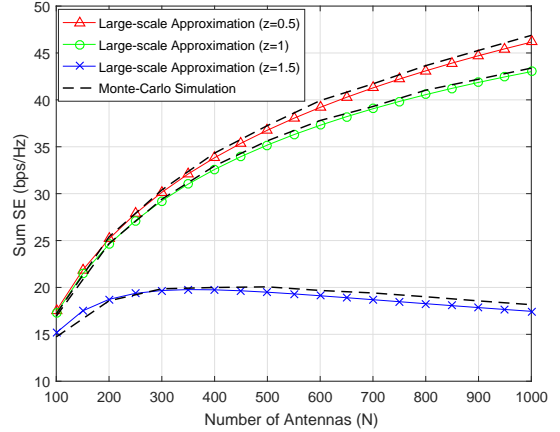}
\caption{Sum SE for different values of $z$ ($\kappa_0=0.05$).}
\label{Fig.5}
\end{figure}

We set $P_u=10$, $P_R=40$, $K=10$, $ \sigma_{{R}}^2=\sigma _{{A_i}}^2=\sigma _{{B_i}}^2=1$, for $i = 1,\ldots,K $ in the Monte Carlo simulation. Moreover, the large-scale fading coefficients $\sigma _{{g_i}}^2$ and $\sigma _{{h_i}}^2$ are arbitrarily generated by i.i.d. ${\cal{N}}({ {1}}, 0.2)$ random variables.
The simulation of $R_{A_i}$ and the large-scale approximation in \eqref{An approximate ergodic achievable rate} are plotted in Fig. \ref{Fig.3}. The simulation results validate the tightness of our large-scale approximation in \eqref{ergodic achievable rate}.\footnote{Note that the analytical results plotted in Figs. 1 and 2 are not below the simulated curves. This is because we consider large-scale approximations that neglect lower order terms that go asymptotically to zero, but might improve the performance when $N$ is not large enough \cite{jin2015ergodic}.} It is clear that the average SE is an increasing function of $N$. Furthermore, when the level of hardware impairments $\kappa$ increases, the average SE decreases. However, high SE of massive MIMO two-way relaying systems with hardware impairments can still be achieved by increasing $N$.

Fig. \ref{Fig.5} validates the hardware scaling law established by Corollary \ref{coro:1}.
Small SE loss can be found when the hardware scaling law is fulfilled ($z= 0.75$), while the curve goes asymptotically to zero when the law is not satisfied ($z=1.5$). When $N$ is small ($N<300$), the gain provided by the massive MIMO system is dominant, while the level of hardware impairments is too small to influence the SE. Therefore the SE will increase with $N$. However, when $N>300$, the loss caused by hardware impairments will exceed the gain of massive MIMO systems so that SE will decrease with $N$.

\section{Conclusion}\label{se:Conclusion}
This letter investigated a multipair two-way relaying system where the relay has a large number of antennas and hardware impairments. The mathematical relationship between the SE, the number of antennas $N$, the number of device pairs $K$, the level of transceiver hardware impairments $\kappa$ has been revealed. Furthermore, a useful hardware scaling law was established to prove that the level of hardware impairments can be increased with ${N}$ without significant SE loss. We conclude that multipair massive MIMO two-way relaying systems can deploy many low-cost transceivers that suffer from hardware impairments.



\appendix
\section{Proof Of Lemma 1}\label{sec:proof}


From \eqref{ergodic achievable rate low bound}, ${\mathbb{E}}\left\{ {{{\left[ {{\text{SINR}}_{{A_i}}} \right]}^{ - 1}}} \right\}$ can be written as
\begin{align}\label{expression of EAi 1}
&{\mathbb{E}}\left\{ {{\frac{1}{  {{\text{SINR}}_{{A_i}}} } }} \right\} \!=\! {{\mathbb{E}}\left\{ {\frac{{{{\left\| {{\bf{g}}_i^T{\bf{F}}}{\boldsymbol{\eta}}_r \right\|}^2}}}{{{{\left\| {{\bf{g}}_i^T{\bf{F}}{{\bf{h}}_i}} \right\|}^2}}} } \right\} \!+\! \frac{1}{{{P_U}{\rho ^2}}}{{\mathbb{E}}\left\{ {\frac{{{{\left\| {{{\bf{g}}_i}{\boldsymbol{\eta}}_t^T} \right\|}^2}}}{{{{\left| {{\bf{g}}_i^T{\bf{F}}{{\bf{h}}_i}} \right|}^2}}}} \right\}}}  \notag \\
&+ \frac{{{\sigma _R}^2}}{{{P_U}}}{\mathbb{E}}\left\{ {\frac{{{{\left\| {{\bf{g}}_i^T{\bf{F}}} \right\|}^2}}}{{{{\left| {{\bf{g}}_i^T{\bf{F}}{{\bf{h}}_i}} \right|}^2}}}} \right\} + \frac{{\sigma _{Ai}^2}}{{{P_U}}}{\mathbb{E}}\left\{ {\frac{1}{{\rho ^2}{{{\left| {{\bf{g}}_i^T{\bf{F}}{{\bf{h}}_i}} \right|}^2}}}} \right\} \notag \\
&+  \sum\limits_{j \ne i}^K {\left( {{\mathbb{E}}\left\{ {\frac{{{{\left| {{\bf{g}}_i^T{\bf{F}}{{\bf{h}}_j}} \right|}^2}}}{{{{\left| {{\bf{g}}_i^T{\bf{F}}{{\bf{h}}_i}} \right|}^2}}}} \right\} + {\mathbb{E}}\left\{ {\frac{{{{\left| {{\bf{g}}_i^T{\bf{F}}{{\bf{g}}_j}} \right|}^2}}}{{{{\left| {{\bf{g}}_i^T{\bf{F}}{{\bf{h}}_i}} \right|}^2}}}} \right\}} \right)} .
\end{align}
Using the law of large numbers \cite[Chapter 3]{cramer2004random}, we have
\begin{align}
\frac{1}{N}{\bf{g}}_i^T{\bf{F}}{{\bf{h}}_i} & - \frac{1}{N}{\left\| {{\bf{g}}_i^*} \right\|^2}{\left\| {{{\bf{h}}_i}} \right\|^2} \xrightarrow[{N \to \infty }]{}  0,\notag\\
\frac{1}{N}{\bf{g}}_i^T{\bf{F}}{{\bf{h}}_j} &- \frac{1}{N}\left( {\left\| {{\bf{g}}_i^*} \right\|^2}{\bf{h}}_i^H{{\bf{h}}_j} + {\left\| {{{\bf{h}}_j}} \right\|^2}{\bf{g}}_i^T{\bf{g}}_j^*\right) \xrightarrow[{N \to \infty }]{}  0,\notag\\
\frac{1}{N}{\bf{g}}_i^T{\bf{F}}{{\bf{g}}_j} &-  \frac{1}{N}\left({\left\| {{\bf{g}}_i^*} \right\|^2}{\bf{h}}_i^H{{\bf{g}}_j} + {\left\| {{{\bf{g}}_j}} \right\|^2}{\bf{g}}_i^T{\bf{h}}_j^*\right) \xrightarrow[{N \to \infty }]{}  0,\notag \\
\frac{1}{N}{\left\| {{\bf{g}}_i^T{\bf{F}}} \right\|^2} &- \frac{1}{N}{\left\| {{\bf{g}}_i^*} \right\|^4}{\left\| {{{\bf{h}}_i}} \right\|^2} \xrightarrow[{N \to \infty }]{}  0.\notag
\end{align}
As $N \to \infty$, we further have
\begin{align}
&{\mathbb{E}}\left\{ {{{\left| {\frac{{{\bf{h}}_i^H{{\bf{h}}_j}}}{{{{\left\| {{{\bf{h}}_i}} \right\|}^2}}} \!+\! \frac{{{{\left\| {{{\bf{h}}_j}} \right\|}^2}}}{{{{\left\| {{{\bf{h}}_i}} \right\|}^2}}}\frac{{{\bf{g}}_i^T{\bf{g}}_j^*}}{{{{\left\| {{\bf{g}}_i^*} \right\|}^2}}}} \right|}^2}} \right\}\!-\! \frac{1}{N}\left( {\frac{{\sigma_{{h_j}}^2}}{{\sigma_{{h_i}}^2}} \!+\! \frac{{\sigma_{{h_j}}^4\sigma _{{g_j}}^2}}{{\sigma_{{h_i}}^4\sigma_{{g_i}}^2}}} \right) \xrightarrow[{N \to \infty }]{}  0,\notag\\
&{\mathbb{E}}\left\{ {{{\left| {\frac{{{\bf{h}}_i^H{{\bf{g}}_j}}}{{{{\left\| {{{\bf{h}}_i}} \right\|}^2}}} \!+\! \frac{{{{\left\| {{{\bf{g}}_j}} \right\|}^2}}}{{{{\left\| {{{\bf{h}}_i}} \right\|}^2}}}\frac{{{\bf{g}}_i^T{\bf{h}}_j^*}}{{{{\left\| {{\bf{g}}_i^*} \right\|}^2}}}} \right|}^2}} \right\}\!-\! \frac{1}{N}\left( {\frac{{\sigma_{{g_j}}^2}}{{\sigma_{{h_i}}^2}} + \frac{{\sigma_{{g_j}}^4\sigma_{{h_j}}^2}}{{\sigma _{{h_i}}^4\sigma_{{g_i}}^2}}} \right)\xrightarrow[{N \to \infty }]{}  0,\notag
\end{align}
\begin{align}\label{C1}
&{\mathbb{E}}\left\{ { {1/{{{\left\| {{{\bf{h}}_i}} \right\|}^2}}}} \right\} -  {1/N{\sigma_{{h_i}}^2}} \xrightarrow[{N \to \infty }]{}  0,\notag\\
&{\mathbb{E}}\left\{ {\frac{{{{\left\| {{\bf{g}}_i^T{\bf{F}}}{\boldsymbol{\eta}}_r \right\|}^2}}}{{{{\left\| {{\bf{g}}_i^T{\bf{F}}{{\bf{h}}_i}} \right\|}^2}}} } \right\} - \frac{P_U \kappa^2 {\sum\limits_{j = 1}^K {\left( {\sigma _{{h_j}}^2 + \sigma_{{{\rm{g}}_j}}^2} \right)} }}{N{\sigma_{{h_i}}^2}} \xrightarrow[{N \to \infty }]{}  0,\notag\\
&{{\mathbb{E}}\left\{ {\frac{{{{\left\| {{{\bf{g}}_i}{\boldsymbol{\eta}}_t^T} \right\|}^2}}}{{{{\left| {{\bf{g}}_i^T{\bf{F}}{{\bf{h}}_i}} \right|}^2}}}} \right\}} - \frac{\kappa^2 P_R}{ N^4 \sigma_{{{\rm{g}}_i}}^2 \sigma _{{{\rm{h}}_i}}^4 } \xrightarrow[{N \to \infty }]{}  0,\notag\\
&{\mathbb{E}}\left\{ { {1/{{{\left\| {{\bf{g}}_i^*} \right\|}^2}{{\left\| {{{\bf{h}}_i}} \right\|}^4}}}} \right\}- {1/N^3{\sigma_{{{\rm{g}}_i}}^2\sigma_{{h_i}}^4}} \xrightarrow[{N \to \infty }]{}  0.
\end{align}
We also have the following properties
\begin{align}\label{rho ^2}
{\mathbb{E}}\left\{{\left\| {\bf{F}}\right\|^2}\right\} &- 2{N^2}\sum\limits_{i = 1}^K {\left( {\sigma_{{g_i}}^2\sigma _{{h_i}}^2} \right)} \xrightarrow[{N \to \infty }]{}  0,\notag \\
{\mathbb{E}}\left\{{\left\| {\bf{FA}} \right\|^2} \right\} &-  {N^3}\sum\limits_{i = 1}^K {\sigma_{{g_i}}^2\sigma _{{h_i}}^2\left( {\sigma_{{g_i}}^2 \!+\! \sigma_{{h_i}}^2} \right)} \xrightarrow[{N \to \infty }]{}  0,\notag\\
{\mathbb{E}}\left\{{\rho ^2}\right\}&- {P_R}/{{{P_U}{N^3}J }}\xrightarrow[{N \to \infty }]{}  0.
\end{align}
By inserting \eqref{C1}-\eqref{rho ^2} into \eqref{expression of EAi 1}, we can complete the proof.

%

\vspace{-0.5cm}
\bibliographystyle{IEEEtran}
\bibliography{IEEEabrv,Ref}

\end{document}